\documentclass[journal,twoside,web]{ieeecolor}
\usepackage{generic}
\usepackage{cite}
\usepackage{amsmath,amssymb,amsfonts}
\usepackage{algorithmic}
\usepackage{graphicx}
\usepackage{algorithm,algorithmic}
\usepackage{hyperref}
\usepackage{booktabs}
\usepackage{multirow}
\usepackage{bm}
\usepackage[table]{xcolor}
\hypersetup{hidelinks=true}
\usepackage{textcomp}
\def\BibTeX{{\rm B\kern-.05em{\sc i\kern-.025em b}\kern-.08em
    T\kern-.1667em\lower.7ex\hbox{E}\kern-.125emX}}
\begin{document}
\title{Prompt-Guided Dual-Path UNet with Mamba for Medical Image Segmentation}
\author{Shaolei Zhang, Jinyan Liu, Tianyi Qian, and Xuesong Li, \IEEEmembership{Member, IEEE}
\thanks{This work was supported in part by the National Natural Science Foundation
of China under Grant 62071049 and Grant 62102026 and in part by Beijing Municipal Natural Science Foundation under Project 4222018. (Corresponding Author: Jinyan Liu and Tianyi Qian.) }
\thanks{Shaolei Zhang, Jinyan Liu, and Xuesong Li are with the School of Computer Science $\&$ Technology, Beijing Institute of Technology, Beijing 100081, China (e-mail: zhangshaolei@bit.edu.cn; jyliu@bit.edu.cn; lixuesong@bit.edu.cn).}
\thanks{Tianyi Qian is with the Qiyuan Laboratory, Beijing 100081, China (e-mail: qiantianyi@qiyuanlab.com).}
}

\maketitle

\begin{abstract}
Convolutional neural networks (CNNs) and transformers are widely employed in constructing UNet architectures for medical image segmentation tasks. However, CNNs struggle to model long-range dependencies, while transformers suffer from quadratic computational complexity. Recently, Mamba, a type of State Space Models, has gained attention for its exceptional ability to model long-range interactions while maintaining linear computational complexity. Despite the emergence of several Mamba-based methods, they still present the following limitations: first, their network designs generally lack perceptual capabilities for the original input data; second, they primarily focus on capturing global information, while often neglecting local details. To address these challenges, we propose a prompt-guided CNN-Mamba dual-path UNet, termed PGM-UNet, for medical image segmentation. Specifically, we introduce a prompt-guided residual Mamba module that adaptively extracts dynamic visual prompts from the original input data, effectively guiding Mamba in capturing global information. Additionally, we design a local-global information fusion network, comprising a local information extraction module, a prompt-guided residual Mamba module, and a multi-focus attention fusion module, which effectively integrates local and global information. Furthermore, inspired by Kolmogorov-Arnold Networks (KANs), we develop a multi-scale information extraction module to capture richer contextual information without altering the resolution. We conduct extensive experiments on the ISIC-2017, ISIC-2018, DIAS, and DRIVE. The results demonstrate that the proposed method significantly outperforms state-of-the-art approaches in multiple medical image segmentation tasks.
\end{abstract}


\section{Introduction}
\label{sec:introduction}
\begin{figure}[!t]
\vspace{0.3cm}
\centering
\centerline{\includegraphics[width=1.0\linewidth]{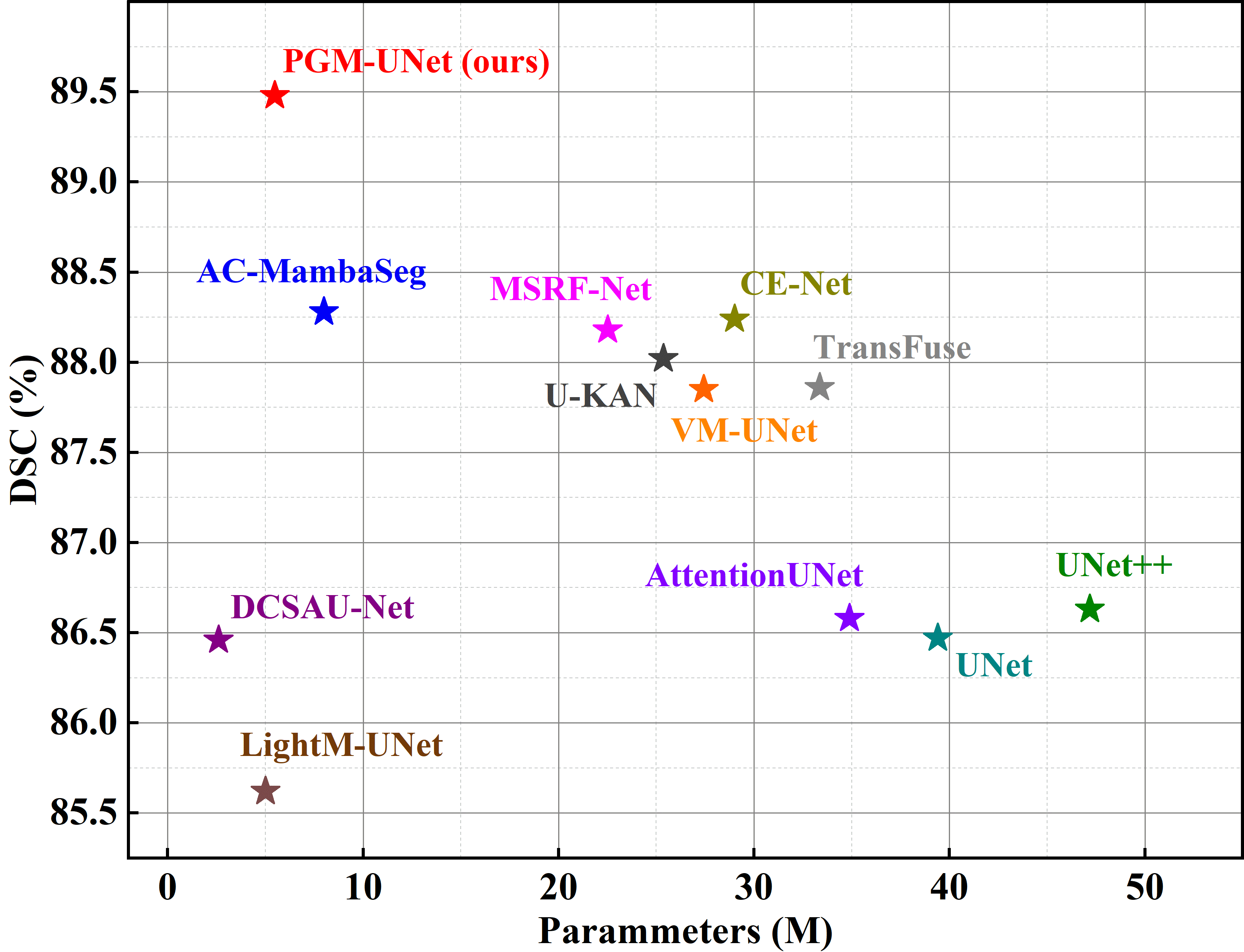}}
\caption{{Visualization of comparison results on the ISIC-2018. The X-axis represents the number of parameters (lower is better), while the Y-axis indicates the DSC values (higher is better). The proposed strategy achieves a good trade-off between segmentation accuracy and parameter efficiency.}}
\label{Fig:1}
\vspace{-0.5cm}
\end{figure}
\IEEEPARstart{M}{edical} image segmentation is a critical technology in clinical analysis, widely applied in various fields such as the skin lesion segmentation~\cite{Ruan2024VMUNetVM, 978-981-97-5128-0, 10733833}, digital subtraction angiography (DSA) segmentation~\cite{LIU2024103247, MENG2020123}, and retinal vessel segmentation~\cite{9994763, 7055281, AZZOPARDI201546}. By accurately identifying regions of interest, segmentation results aids in lesion assessment, facilitates early disease detection, and enhance treatment planning. Manual pixel-level annotation of medical images is a laborious and costly process that typically requires collaboration among multiple medical experts. In contrast, automated medical image segmentation techniques provide a faster and more reliable diagnostic solution, enhancing diagnostic accuracy and improving the efficiency of patient care. Traditional methods~\cite{932744} typically based on edge detection and template matching often struggle to handle the complexity and variability of pathological features, leading to suboptimal results and limiting their clinical applicability.

With the advancement of deep learning (DL) in the field of medical image processing, DL-based methods have become the mainstream in the field of medical image segmentation. In recent years, convolutional neural networks (CNNs) have effectively overcome the limitations of traditional segmentation techniques. Ronneberger et al. \cite{10.1007/978-3-319-24574-4_28} introduced U-Net, an encoder-decoder framework that enhances feature extraction through skip connections. Building on this, Oktay et al. \cite{Oktay2018AttentionUL} incorporated attention mechanisms to improve feature selection, while Zhou et al. \cite{10.1007/978-3-030-00889-5_1} refined the architecture with dense skip connections. To improve efficiency for real-world applications, Ruan et al. \cite{9995040} reduced model complexity by optimizing channel configurations and integrating multiple attention modules while maintaining strong segmentation performance. Despite these advancements, CNN-based models struggle to capture long-range dependencies, limiting segmentation accuracy in complex medical images. 
Transformers provide a potential solution by leveraging a multi-head self-attention mechanism to effectively model global dependencies. Chen et al. \cite{Chen2021TransUNetTM} pioneered the integration of the vision transformer (ViT) \cite{Dosovitskiy2020AnII} into U-Net for medical image segmentation, while Cao et al. \cite{10.1007/978-3-031-25066-8_9} introduced the swin transformer \cite{Liu2021SwinTH} to further enhance segmentation performance. However, although transformers excel at capturing global contextual information, they are less effective at preserving fine-grained local details. To address this limitation, Zhang et al. \cite{10.1007/978-3-030-87193-2_2} proposed a dual-path structure that combines CNNs and ViT to capture local and global information, respectively. While transformer-based models improve global information modeling, their self-attention mechanism introduces high computational complexity, posing challenges for medical image segmentation.

Recently, State Space Models (SSMs) \cite{Gu2021EfficientlyML} have gained significant attention in medical image segmentation research. Based on classical SSM research, Mamba \cite{Gu2023MambaLS} stands out for its ability to model long-range dependencies while maintaining linear computational complexity, making it a promising alternative to transformers. Several studies have explored the application of Mamba-based models in medical image segmentation. Ma et al. \cite{Ma2024UMambaEL} introduced a novel CNN-SSM hybrid model, marking its first application to medical image segmentation. Xing et al. \cite{978-3-031-72111-3} incorporated SSMs in the encoder part, while still using CNNs in the decoder part. Inspired by the success of VMamba \cite{Liu2024VMambaVS} in image classification tasks, Ruan et al. \cite{Ruan2024VMUNetVM} proposed a pure SSM-based model, aiming to demonstrate the potential of SSMs in medical image segmentation. Liao et al. \cite{Liao2024LightMUNetMA} developed a light-weight UNet based on Mamba, which significantly reduced the number of parameters but suffered from insufficient performance. To improve the performance of skin lesion segmentation, Nguyen et al. \cite{10.1007/978-3-031-76197-3_2} introduced a UNet with a hybrid CNN-Mamba backbone, demonstrating impressive performance and achieving favorable results. Nevertheless, these Mamba-based methods neglect the importance of local details, thereby limiting their performance. Moreover, the aforementioned methods generally lack the dynamic perception capability of the original input data. To address these challenges, we propose a prompt-guided CNN-Mamba dual-path UNet termed PGM-UNet for medical image segmentation. The main contributions can be summarized as follows:

\begin{itemize}
	\item We propose a prompt-guided residual Mamba module that adaptively extracts dynamic visual prompts from the original input data, thereby guiding Mamba to effectively capture global information. Specifically, the visual prompts optimize the information extraction process at each stage of the UNet architecture, significantly enhancing the model's ability to perceive and process diverse original input data. The prompt-guided residual Mamba module can be regarded as a plug-and-play module for global information extraction, and integrated into other image processing tasks. 
	
	\item We come up with a local-global information fusion network consisting of a local information extraction module, a prompt-guided residual Mamba module, and a multi-focus attention fusion module. This network effectively captures both low-level spatial features and high-level semantic context. The multi-focus attention fusion module integrates multi-level features from the local information extraction module and the prompt-guided residual Mamba module using channel attention mechanisms. To the best of our knowledge, the proposed method is the first parallel branch model that combines CNNs with Mamba.
	
	\item We develop a multi-scale information extraction module that integrates multiple dilated convolutions with Kolmogorov-Arnold Networks (KANs). By employing dilated convolutions of different sizes, the module can effectively capture contextual information while maintaining resolution. Furthermore, the module leverages the powerful nonlinear representation capability of the KANs to deeply mine deep features, significantly enhancing the interpretability of the model. In light of the KANs exceptional performance in high-dimensional complex functions, we adopt the multi-scale information extraction module as the bottleneck layer of the proposed framwork. As illustrated in Fig.~\ref{Fig:1}, the proposed framework achieves a good balance between segmentation performance and computational efficiency. Experimental results demonstrate that our method outperforms state-of-the-art (SOTA) methods across various tasks.
	
\end{itemize}
\begin{figure*}[!t]
\centering
\vspace{-0.3cm}
\centerline{\includegraphics[width=18cm]{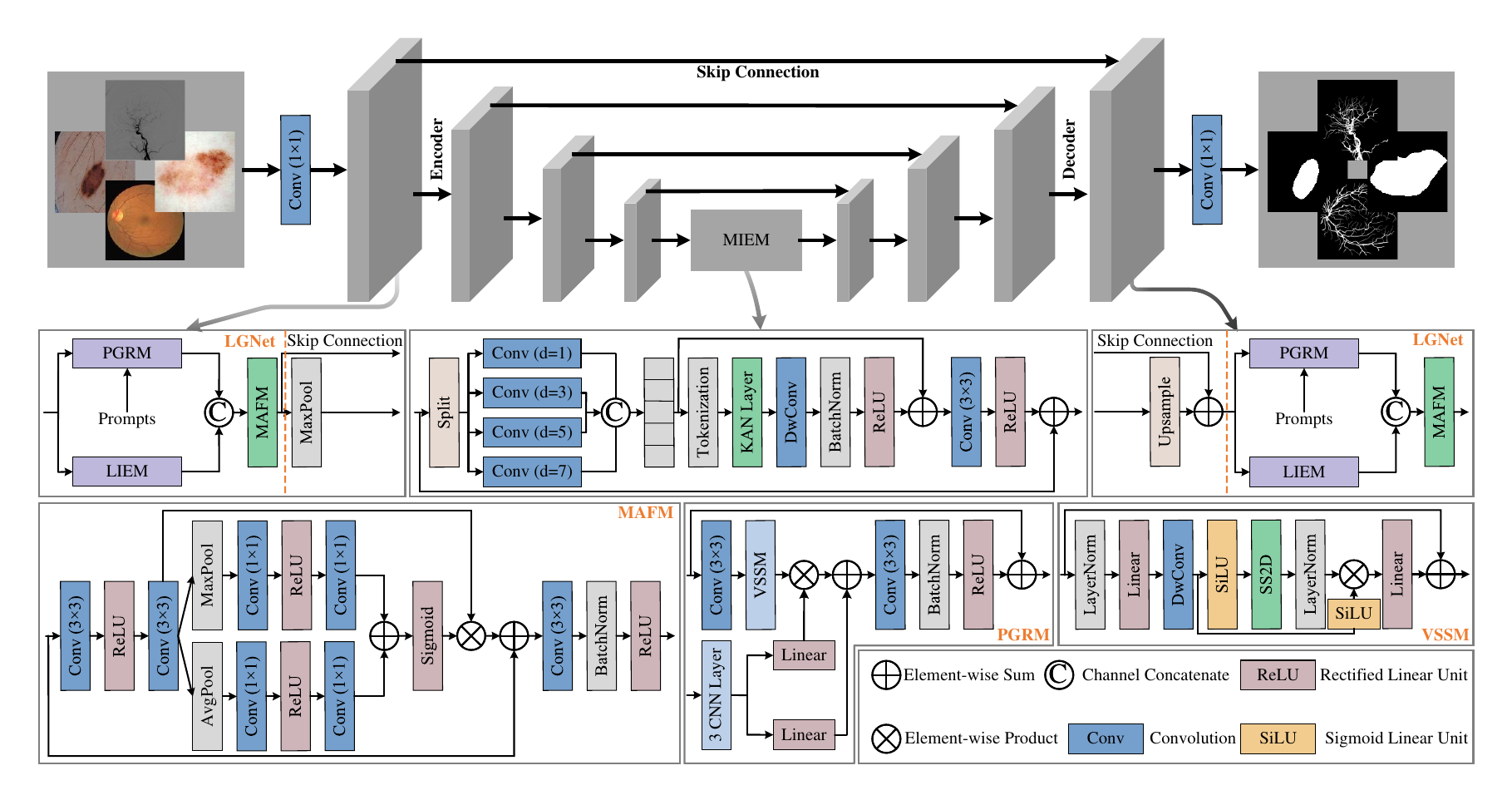}}
\caption{{Overview of the proposed PGM-UNet architecture. The encoder and decoder are primarily composed of the LG-Net. The LIEM is used to extract local information, while the PGRM extracts global information under the guidance of prompt information. The MAFM employs channel attention mechanisms to reweight the local information from LIEM and the global information from PGRM, enabling effective integration. Additionally, the MIEM is constructed by combining multiple dilated convolutions and KANs, which serves as the bottleneck layer.}}
\label{Fig:2}
\vspace{-0.3cm}
\end{figure*}
\section{Methods}
Fig.~\ref{Fig:2} presents the overall architecture of the proposed PGM-UNet, comprising an input convolutional layer, an encoder, a decoder, a bottleneck layer, skip connections, and an output convolutional layer. In the following sections, we will provide a detailed description of each component and its role in the segmentation process.
\subsection{Local-Global Information Fusion Network}
Although Mamba demonstrates exceptional capability in modeling global context, it still faces challenges in capturing local details. To address this limitation, recent studies have attempted to integrate CNNs with Mamba architectures in a sequential manner~\cite{Ma2024UMambaEL, Liao2024LightMUNetMA, 10.1007/978-3-031-76197-3_2} for medical image segmentation. However, these approaches face challenges in effectively combining local and global information, which leaves room for further improvement. To fully leverage the strengths of both CNNs and Mamba, we propose a dual-path parallel integration strategy, forming the local-global information fusion network termed LG-Net. This dual-path CNN-Mamba architecture enables the effective fusion of fine-grained local details with comprehensive global context, thereby enhancing segmentation performance.

As illustrated in Fig.~\ref{Fig:2}, LG-Net consists of three key components: the local information extraction module (LIEM), the prompt-guided residual Mamba module (PGRM), and the multi-focus attention fusion module (MAFM). Specifically, LIEM adopts a residual structure, i.e., [Conv + BN + ReLU+ $\dots$ + Conv + BN + Skip Connection], to effectively extract local information. MAFM employs residual channel attention mechanisms \cite{10.1007/978-3-030-01234-2_18} and CNNs to adaptively fuse the local information from LIEM and the global information from PGRM. MAFM compresses spatial information using both max pooling and average pooling, which enhances the relevant channels while mitigating performance degradation caused by redundant information. By leveraging MAFM, LG-Net can more effectively focus on the key regions of the feature while suppressing irrelevant details. Mathematically, the process of LG-Net can be formulated as 
\begin{align}
	{\bm f}_{\rm out} = {\rm MAFM}[{\rm LIEM}({\bm f}_{\rm in}),{\rm PGRM}({\bm f}_{\rm in})]
	\label{eq:4}
\end{align}
where [${\bm a}$, ${\bm b}$] denotes the channel-wise concatenation operation of ${\bm a}$ and ${\bm b}$.

Recently, prompt learning strategies \cite{potlapalli2023promptir, 10833823} have been shown to significantly enhance the performance of CNNs and transformers. Notably, adaptive prompt learning \cite{Kong2024TowardsEM} stands out, as it enables the network to autonomously determine which features to extract, thereby improving the network's generalization capabilities. Based on this, we have attempted to combine adaptive prompt learning with Mamba to construct a prompt-guided residual Mamba module. Although prompt learning has been widely applied in constructing all-in-one models, the imbalance in data distribution across different medical image segmentation tasks poses a challenge to building a all-in-one model. In this paper, we focus exclusively on enhancing the performance of a single medical image segmentation task.
Specifically, we take the original input image as the visual prompt ${\bm p}$ and use 3-CNN layers \cite{Kong2024TowardsEM} as the feature extractor ${\rm F}_{\rm ext}(\cdot)$ to extract features from the ${\bm p}$. Subsequently, by applying fully connected layers ${\rm FC}(\cdot)$, we transform these extracted features into the appropriate shape (1$\times$channel or dimension) suitable for the corresponding module's output, which includes determining the weight $k$ and the bias $b$. Finally, we multiply the weight $k$ and add the bias $b$ to the output features. Mathematically, the process of PGRM can be represented as
\begin{equation}
	k = {\rm FC}_1({\rm F}_{\rm ext}({\bm p})), b = {\rm FC}_2({\rm F}_{\rm ext}({\bm p})),
	\label{eq:5}
\end{equation}
\begin{equation}
	{\bm f}_{\rm global} = {\rm VSSM}({\rm Conv}({\bm f}_{\rm in})) \ast k + b,
	\label{eq:6}
\end{equation}
and
\begin{equation}
	{\bm f}_{\rm global} = {\rm ReLU}({\rm BN}({\rm Conv}({\bm f}_{\rm global}))) + {\bm f}_{\rm in}
	\label{eq:7}
\end{equation}
where ${\bm f}_{\rm global}$ is the output feature of the PGRM. The prompt features are used to guide the Mamba structure to effectively extract global information.
\subsection{Encoder and Decoder}
As shown in Fig.~\ref{Fig:2}, the encoder consists of four stages with channel dimensions of [8, 16, 32, 64], respectively. At each stage of the encoder, LG-Net is utilized to efficiently extract both local and global information while progressively increasing the number of channels. Before the resolution is reduced through max pooling, the feature maps are leveraged for skip connections. Max pooling with a stride of two is then used to downsample the resolution. Similarly, the decoder also consists of four stages. At each stage of the decoder, we upsample the features from either the bottleneck layer or the previous stage and combine them with the skip connection information. Subsequently, LG-Net is employed to efficiently fuse the extracted information and reduce the number of channels.
\subsection{Bottleneck Layer}
Fig.~\ref{Fig:2} depicts the structure of the bottleneck layer, i.e., the multi-scale information extraction module (MIEM). As a critical component of PGM-UNet, the bottleneck layer serves as the bridge between the encoder and decoder. Effectively extracting and utilizing information from the encoder remains a key challenge in improving segmentation performance.

Inspired by the Kolmogorov-Arnold representation theorem \cite{1957On}, Liu et al. \cite{Liu2024KANKN} proposed Kolmogorov-Arnold Networks, which replace traditional fixed activation functions with learnable functions, thereby achieving stronger nonlinear expressiveness and higher interpretability. Moreover, KANs have demonstrated outstanding performance in approximating high-dimensional complex functions and have been successfully applied across various applications \cite{Li2024UKANMS, Li2024HSRKANEH}. In medical image segmentation, the output of the encoder after several downsamplings is expected to capture contextual information. However, further reducing the resolution may compromise the structural consistency of the restored image. 
To capture more contextual information without altering the resolution, we propose a multi-scale information extraction module constructed using several convolutions with different dilation rates and KANs. Multiple dilated convolutions generate receptive fields of varying sizes, thereby facilitating the smooth extraction of multi-scale information. To reduce the computational cost of multiple convolutions, the MIEM compresses the channels of each convolution, ensuring that the total number of concatenated channels matches the output channels. This process can be represented as
\begin{equation}
	\hat{{\bm f}} = {\rm Concat}({\rm Conv}_{i}^{d}({\bm f}_{\rm en})|i,d\in \left\lbrace1,2,3,4\right\rbrace),
	\label{eq:8}
\end{equation}
\begin{equation}
	{\bm f}_{\rm kan} = {\rm ReLU}({\rm BN}({\rm DwConv}({\rm KAN}(\hat{{\bm f}}))))) + \hat{{\bm f}},
	\label{eq:9}
\end{equation}
and
\begin{equation}
	{\bm f}_{\rm miem} = {\rm ReLU}({\rm Conv}({\bm f}_{\rm kan}))) + {\bm f}_{\rm en}
	\label{eq:10}
\end{equation}
where ${\rm Conv}_{i}^{d}(\cdot)$ is the $i$-th convolution with the dilation rate $d$, and ${\rm Concat}(\cdot)$ denotes the channel-wise concatenation operation.
\subsection{Loss Function}
In this paper, we employ binary cross-entropy (BCE) and Dice loss for supervision to generate more accurate mask information. The total loss function can be represented as
\begin{equation}
	\mathcal{L}_{\rm total} = \mathcal{L}_{\rm BCE} + \lambda \mathcal{L}_{\rm Dice},
	\label{eq:11}
\end{equation}
\begin{equation}
	\mathcal{L}_{\rm BCE} = -\frac{1}{N} \sum_{i=1}^{N} \left[ y_{i}{\rm log} (\hat{y}_{i}) + (1-y_{i}){\rm log}(1-\hat{y}_{i})\right],
	\label{eq:12}
\end{equation}
\begin{equation}
	\mathcal{L}_{\rm Dice} = 1 - \frac{2|X \cap Y|}{|X| + |Y|}
	\label{eq:13}
\end{equation}
where $\lambda$ is a hyper-parameter that balances the weights of different loss components. $N$ denotes the total number of samples. $y_i$ and $\hat{y}_{i}$ signify the ground truth and prediction, respectively. $|X|$ and $|Y|$ represent the ground truth and prediction, respectively.
\section{Results}
\begin{table*}[!t]
\vspace{-0.3cm}
\centering
\caption{Quantitative comparisons on the ISIC-2017 and ISIC-2018. The best and second best values are highlighted in {\bf Bold} and \underline{underline}.}
\renewcommand{\arraystretch}{1}
{\footnotesize{\tabcolsep=4pt\begin{tabular}{c|c|c|cc|cc|cc}
\hline \hline
\multirow{2}{*}{\bf Methods} & \multirow{2}{*}{\bf Journal/Conference} & \multirow{2}{*}{\bf Parameters} & \multicolumn{2}{c|}{\bf ISIC-2017} & \multicolumn{2}{c|}{\bf ISIC-2018} & \multicolumn{2}{c}{\bf Average} \\ \cline{4-9} 
& & & \bf DSC (\%) $\textcolor{red}{\uparrow}$ & \bf mIoU (\%) $\textcolor{red}{\uparrow}$ & \bf DSC (\%) $\textcolor{red}{\uparrow}$ & \bf mIoU (\%) $\textcolor{red}{\uparrow}$ & \bf DSC (\%) $\textcolor{red}{\uparrow}$& \bf mIoU (\%) $\textcolor{red}{\uparrow}$ \\ 
\hline
\multicolumn{1}{c|}{UNet \cite{10.1007/978-3-319-24574-4_28 }} & MICCAI, 2015 & 39.40M & 82.24 & 73.21 & 86.47 & 78.94 & 84.36 & 76.08 \\ 
\multicolumn{1}{c|}{UNet++ \cite{10.1007/978-3-030-00889-5_1}} & DLMIA, 2018 & 47.18M & 82.17 & 73.07 & 86.63 & 79.04 & 84.40 & 76.06 \\ 
\multicolumn{1}{c|}{AttentionUNet \cite{Oktay2018AttentionUL}} & arXiv, 2018 & 34.88M & 83.25 & 74.55 & 86.58  & 78.78  & 84.92 & 76.67 \\  
\multicolumn{1}{c|}{CE-Net \cite{8662594}} & TMI, 2019 & 29.00M & 83.95 & 75.55 &88.24  &80.61  & 86.09 & 78.08 \\ 
\multicolumn{1}{c|}{TransFuse \cite{10.1007/978-3-030-87193-2_2}} & MICCAI, 2021 & 33.36M & 83.00 & 74.11 &87.86  &80.34  & 85.43 & 77.23 \\ 
\multicolumn{1}{c|}{MSRF-Net \cite{9662196}} & JBHI, 2021 & 22.51M & 84.04 & 75.67 & 88.18  & \underline{80.82}  & 86.11 & 78.25 \\ 
\multicolumn{1}{c|}{MALUNet \cite{9995040}} & MALUNet & 0.18M & 82.82 & 73.90 & 84.29  & 75.44 & 83.55 & 74.67 \\ 
\multicolumn{1}{c|}{DCSAU-Net \cite{XU2023106626}} & CBM, 2023 & 2.60M & 82.49  & 73.83 & 86.46  & 78.50  & 84.48 & 76.17 \\ 
\multicolumn{1}{c|}{LightM-UNet \cite{Liao2024LightMUNetMA}} & arXiv, 2024 & 5.01M &80.11  &71.46  &85.62  &77.25  & 82.86 & 74.36 \\ 
\multicolumn{1}{c|}{U-KAN \cite{Li2024UKANMS}} & arXiv, 2024 & 25.36M &83.57  &75.06  &88.02  &80.61  & 85.80 & 77.84 \\
\multicolumn{1}{c|}{VM-UNet \cite{Ruan2024VMUNetVM}} & arXiv, 2024 & 27.43M &83.50  &75.22  &87.85  & 80.15  & 85.68 & 77.69 \\ 
\multicolumn{1}{c|}{AC-MambaSeg \cite{10.1007/978-3-031-76197-3_2}} & arXiv, 2024 & 7.99M & \underline{84.78}  & \underline{76.42} & \underline{88.28} & {80.70} & \underline{86.53} & \underline{78.56} \\
\rowcolor{gray!25}
\multicolumn{1}{c|}{PGM-UNet} & Ours & 5.48M & {\bf 85.89}  & {\bf 77.77}  & {\bf 89.48}  & {\bf 82.68}  & {\bf 87.69} & {\bf 80.23} \\ \hline \hline
\end{tabular}}}
\label{tab:1}
\vspace{-0.3cm}
\end{table*}
\begin{figure*}[htbp]
\centering
{\includegraphics[width=1.0\linewidth]{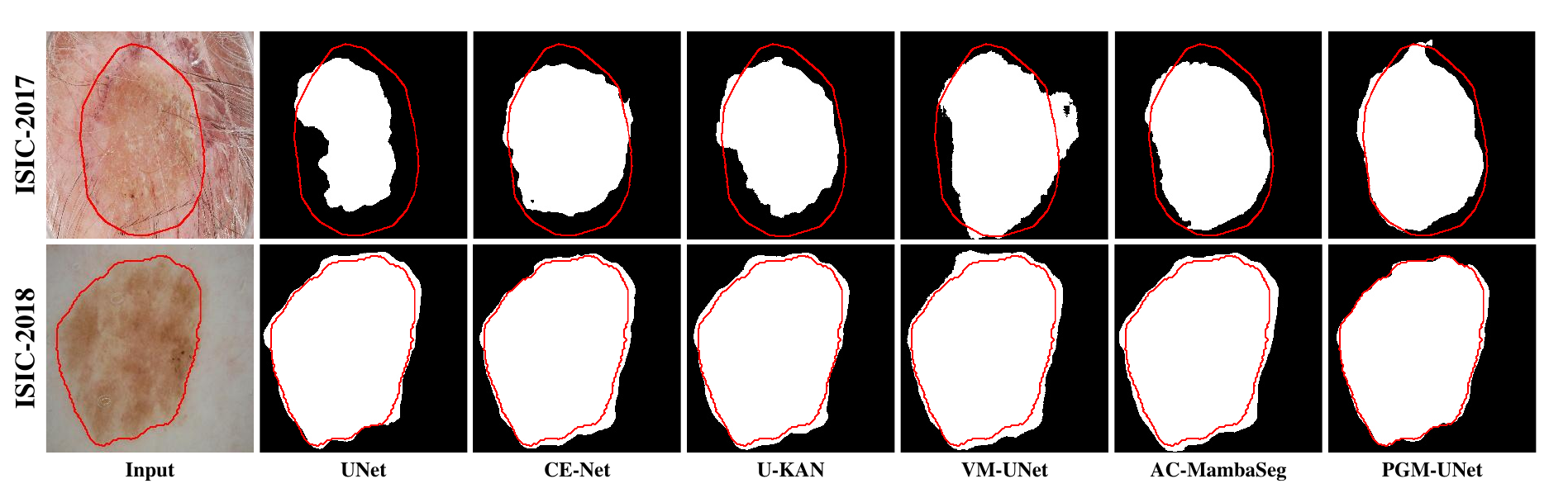}}
\caption{{Visual comparison of different medical image segmentation approaches on skin lesion images selected from the ISIC-2017 and ISIC-2018. Red lines indicate the boundaries of the labels.}}
\label{Fig:3}
\end{figure*}
\begin{table*}[htbp]
\vspace{-0.3cm}
\centering
\caption{Quantitative comparisons on the DIAS. The best and second best values are highlighted in {\bf Bold} and \underline{underline}.}
\renewcommand{\arraystretch}{1}
{\footnotesize{\tabcolsep=8pt\begin{tabular}{c|c|c|c|c|c|c|c} 
\hline
\hline
{\bf Methods}  &{\bf Journal/Conference
} &{\bf Parameters} &{\bf DSC (\%) $\textcolor{red}{\uparrow}$} &{\bf mIoU (\%) $\textcolor{red}{\uparrow}$}  &{\bf Acc (\%) $\textcolor{red}{\uparrow}$} &{\bf AUC (\%) $\textcolor{red}{\uparrow}$} &{\bf Sen (\%) $\textcolor{red}{\uparrow}$}\\ 
\hline
{UNet \cite{10.1007/978-3-319-24574-4_28 }} &MICCAI, 2015    &39.40M    &65.07    &48.81    &95.00    &80.02 &62.12\\
{UNet++ \cite{10.1007/978-3-030-00889-5_1}} &DLMIA, 2018   &47.18M   &66.43    &50.33  &95.09    &80.86 &63.85\\
{AttentionUNet \cite{Oktay2018AttentionUL}} &arXiv, 2018   &34.88M  &73.67    &58.60  &96.06  &85.65 &73.19\\
{CE-Net \cite{8662594}} &TMI, 2019   &29.00M   &70.72    &54.86  &95.66    &83.40 &68.82  \\
{MSRF-Net \cite{9662196}} &JBHI, 2021   &22.51M   &67.49    &51.53   &95.63    &79.72 &60.74 \\
{DCSAU-Net \cite{XU2023106626}} &CBM, 2023  &2.60M   &67.64    &51.59    &95.36    &81.70 &65.36  \\
{LightM-UNet \cite{Liao2024LightMUNetMA}} &arXiv, 2024   &5.01M   &73.28    &58.03   &95.89    &\underline{86.26} &\underline{74.73}\\
{U-KAN \cite{Li2024UKANMS}} &arXiv, 2024   &25.36M   &71.41 &56.08  &95.89  &83.36 &68.46  \\
{AC-MambaSeg \cite{10.1007/978-3-031-76197-3_2}} &arXiv, 2024   &7.99M   &\underline{74.19}    &\underline{59.12}   &\underline{96.08}    &85.99 &73.99\\
\rowcolor{gray!25}
PGM-UNet &Ours   &5.48M   &{\bf 75.31}   &{\bf 60.60}   &{\bf 96.20}    &{\bf 87.54} &{\bf 77.21}\\
\hline
\hline
\end{tabular}}}
\label{tab:2}
\vspace{-0.3cm}
\end{table*}
\subsection{Dataset}
To evaluate the performance of the proposed PGM-UNet, we use several publicly available medical imaging datasets: the ISIC-2017 Challenge dataset \cite{8363547}, the ISIC-2018 Challenge dataset \cite{Codella2019SkinLA, tschandl2018ham10000}, DIAS \cite{LIU2024103247}, DRIVE \cite{1282003}, and PH2 \cite{6610779}. Specifically, the ISIC-2017 comprises 2,000 images for training, 150 for validation, and 600 for testing. Similarly, the ISIC-2018 includes 2,594 images for training, 100 for validation, and 1,000 for testing. DIAS consists of 60 DSA images, with 30 allocated for training, 10 for validation, and the remaining 20 for testing. DRIVE contains 40 color retinal images, of which 20 are used for training and the remaining 20 for testing. Additionally, the PH2 includes a total of 200 images, which are employed for generalization experiments to evaluate the performance of the proposed PGM-UNet on unseen data.
\subsection{Implementation Details and Evaluation Criteria}
The proposed PGM-UNet is implemented in PyTorch and trained on a single NVIDIA GeForce RTX 3090 GPU with 24 GB of memory. We set the batch size to $16$ and use Adam optimizer with an initial learning rate of $1\times10^{-4}$. To prevent overfitting, we employ both an early stopping mechanism and the ReduceLROnPlateau learning rate scheduler. In our experiments, we manually set $\lambda=1$. For the ISIC-2017 and ISIC-2018, the images are resized to $256 \times 256$ pixels, while the DIAS and DRIVE images are resized to $512 \times 512$ pixels. Data augmentation techniques, including random rotation, vertical flipping, horizontal flipping, and coarse dropout, are employed to enhance the diversity of the training data. The final segmentation results are obtained by binarizing the predicted probability maps using a threshold of $0.5$. To ensure fairness, all experiments are conducted under consistent experimental conditions across all methods.

To evaluate the performance of our proposed PGM-UNet, we use standard computer vision metrics commonly employed in medical image segmentation, including the dice similarity coefficient (DSC), mean intersection over union (mIoU), accuracy (Acc), sensitivity (Sen), and area under the curve (AUC).
\subsection{Performance Evaluation}
To demonstrate the superiority of PGM-UNet, we conduct both quantitative and visual comparisons with several SOTA medical image segmentation methods, including UNet \cite{10.1007/978-3-319-24574-4_28 }, UNet++ \cite{10.1007/978-3-030-00889-5_1}, AttentionUNet \cite{Oktay2018AttentionUL}, CE-Net \cite{8662594}, MSRF-Net \cite{9662196}, DCSAU-Net \cite{XU2023106626}, LightM-UNet \cite{Liao2024LightMUNetMA}, U-KAN \cite{Li2024UKANMS}, and AC-MambaSeg \cite{10.1007/978-3-031-76197-3_2}.
\subsubsection{Comparison on Skin Lesion Segmentation}
Table~\ref{tab:1} provides a comprehensive comparison of the quantitative performance of various methods on the ISIC-2017 and ISIC-2018. As shown in Table~\ref{tab:1}, PGM-UNet outperforms other SOTA methods across all metrics for both skin lesion segmentation datasets, achieving an average DSC of 87.69$\%$ and an average mIoU of 80.23$\%$. Furthermore, compared to Mamba-based methods such as LightM-UNet, VM-UNet, and AC-MambaSeg, our approach achieves superior performance by effectively integrating the global information provided by Mamba with the local information from CNNs. Specifically, compared to AC-MambaSeg, PGM-UNet improves the average DSC by 1.16$\%$ and the average mIoU by 1.67$\%$. 

Fig.~\ref{Fig:3} presents a visual comparison of PGM-UNet with other SOTA methods on the ISIC-2018. As shown in Fig.~\ref{Fig:3}, PGM-UNet accurately delineates skin lesion boundaries, regardless of their size and shape, further highlighting its effectiveness.
\begin{table*}[htbp]
\vspace{-0.3cm}
\centering
\caption{Quantitative comparisons on the DRIVE. The best and second best values are highlighted in {\bf Bold} and \underline{underline}.}
\renewcommand{\arraystretch}{1}
{\footnotesize{\tabcolsep=8pt\begin{tabular}{c|c|c|c|c|c|c|c} 
\hline
\hline
{\bf Methods}  &{\bf Journal/Conference
} &{\bf Parameters} &{\bf DSC (\%) $\textcolor{red}{\uparrow}$} &{\bf mIoU (\%) $\textcolor{red}{\uparrow}$}  &{\bf Acc (\%) $\textcolor{red}{\uparrow}$} &{\bf AUC (\%) $\textcolor{red}{\uparrow}$} &{\bf Sen (\%) $\textcolor{red}{\uparrow}$}\\ 
\hline
{UNet \cite{10.1007/978-3-319-24574-4_28 }} &MICCAI, 2015    &39.40M    &56.62    &39.91    &93.33    &73.76 &50.01\\
{UNet++ \cite{10.1007/978-3-030-00889-5_1}} &DLMIA, 2018   &47.18M   &71.46    &55.78  &95.41    &82.58 &66.99\\
{AttentionUNet \cite{Oktay2018AttentionUL}} &arXiv, 2018   &34.88M  &\underline{81.40}    &\underline{68.67}  &\underline{96.75}   &\underline{90.13} &\underline{82.09}\\
{CE-Net \cite{8662594}} &TMI, 2019   &29.00M   &78.65    &64.83  &96.26    &88.55 &79.17  \\
{MSRF-Net \cite{9662196}} &JBHI, 2021   &22.51M   &79.42    &65.91   &96.38    &89.25 &80.57 \\
{DCSAU-Net \cite{XU2023106626}} &CBM, 2023  &2.60M   &78.42    &64.56    &96.35    &87.49 &76.72  \\
{LightM-UNet \cite{Liao2024LightMUNetMA}} &arXiv, 2024   &5.01M  &77.98 &63.94   &96.21    &87.70 &77.38\\
{U-KAN \cite{Li2024UKANMS}} &arXiv, 2024   &25.36M   &79.48    &65.97  &96.38  &89.36 &80.83  \\
{AC-MambaSeg \cite{10.1007/978-3-031-76197-3_2}} &arXiv, 2024   &7.99M   &80.04    &66.74   &96.48    &89.56 &81.15\\
\rowcolor{gray!25}
PGM-UNet &Ours   &5.48M   &{\bf 81.81}   &{\bf 69.24}    &{\bf 96.80}    &{\bf 90.51} &{\bf 82.86}\\
\hline
\hline
\end{tabular}}}
\label{tab:3}
\end{table*}
\begin{figure*}[htbp]
\vspace{-0.3cm}
\centering
{\includegraphics[width=1.0\linewidth]{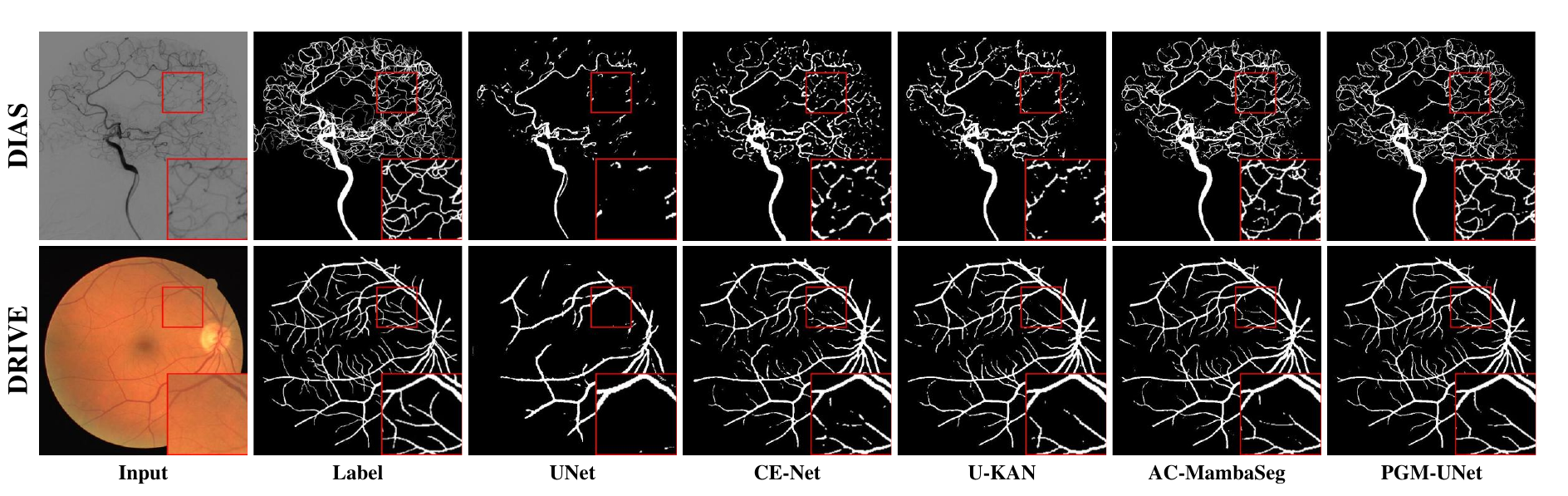}}
\caption{{Visual comparison of different medical image segmentation approaches on vessel segmentation images selected from the DIAS and DRIVE. Zooming in on the red screen enhances the viewing experience.}}
\label{Fig:4}
\vspace{-0.3cm}
\end{figure*}
\begin{table}[htbp]
\centering
\caption{Quantitative comparisons on the PH2 with unseen clinical settings. The best and second best values are highlighted in {\bf Bold} and \underline{underline}.}
\renewcommand{\arraystretch}{1}
{\footnotesize{\tabcolsep=7pt\begin{tabular}{c|c|c|c} 
\hline
\hline
{\bf Methods }&\bf DSC (\%) $\textcolor{red}{\uparrow}$ &\bf mIoU (\%) $\textcolor{red}{\uparrow}$  &\bf Acc (\%) $\textcolor{red}{\uparrow}$\\
\hline
{UNet \cite{10.1007/978-3-319-24574-4_28 }}  &89.53  &82.24   &92.71\\
{UNet++ \cite{10.1007/978-3-030-00889-5_1}} &89.29   &81.93 &92.82  \\ 
{AttentionUNet \cite{Oktay2018AttentionUL}}  &91.04  &84.21 &94.02  \\				
{CE-Net \cite{8662594}} &90.91  &84.17 &94.10   \\
{TransFuse \cite{10.1007/978-3-030-87193-2_2}} &91.75  &85.42 &94.82  \\ 
{MSRF-Net \cite{9662196}} &91.42   &84.75  &94.69  \\ 
{MALUNet \cite{9995040}} &87.84    &79.67  &92.02  \\ 
{DCSAU-Net \cite{XU2023106626}} &91.17   &84.56  &94.53  \\ 
{LightM-UNet \cite{Liao2024LightMUNetMA}}  &91.42  &84.67  &94.63   \\
{U-KAN \cite{Li2024UKANMS}}  & \underline{91.94}  &\underline{85.72}  &\underline{95.02}   \\ 
{VM-UNet \cite{Ruan2024VMUNetVM}}  &91.62   &85.03 &94.98  \\
{AC-MambaSeg \cite{10.1007/978-3-031-76197-3_2}}  &91.75   &85.36 &94.79 \\ 
\rowcolor{gray!25}
{PGM-UNet}  &{\bf 92.01}  &{\bf 85.99}  &{\bf 95.09} \\ 
\hline
\hline
\end{tabular}}}
\vspace{-0.5cm}
\label{tab:4}
\end{table}
\begin{table*}[htbp]
\centering
\caption{Ablation Studies of PGM-UNet on the ISIC-2018. The best and second best values are highlighted in {\bf Bold} and \underline{underline}.}
\renewcommand{\arraystretch}{1}
{\footnotesize{\tabcolsep=10pt\begin{tabular}{c|c|c|c|c|c} 
\hline
\hline
{\bf Methods} &{\bf Parameters} &{\bf DSC (\%) $\textcolor{red}{\uparrow}$} &{\bf mIoU (\%) $\textcolor{red}{\uparrow}$}  &{\bf Acc (\%) $\textcolor{red}{\uparrow}$} &{\bf Sen (\%) $\textcolor{red}{\uparrow}$} \\ 
\hline
PGM-UNet w/o MIEM   &4.19M   &88.07    &81.18  &92.98    &90.23 \\
PGM-UNet w/o LIEM  &2.72M   &88.81    &81.82  &\underline{93.41}    &\underline{92.79}  \\
PGM-UNet w/o PGRM   &3.07M  &88.66    &81.51  &92.97    &92.50 \\
PGM-UNet w/o Prompt   &5.23M    &\underline{88.70}    &\underline{82.01}    &93.11    &90.44 \\
PGM-UNet w/o Attention    &3.50M   &88.34   &81.32  &93.35    &91.94 \\
\rowcolor{gray!25}
PGM-UNet   &5.48M   &{\bf 89.48} &{\bf 82.68}    &{\bf 93.54}    &{\bf 93.65}\\
\hline
\hline
\end{tabular}}}
\label{tab:5}
\vspace{-0.3cm}
\end{table*}
\subsubsection{Comparison on Vessel Segmentation}
The automated segmentation of intracranial arteries in DSA is essential for quantifying vascular morphology, significantly contributing to computer-assisted stroke research and clinical practice. To evaluate the performance of PGM-UNet in vascular segmentation for DSA images, we conduct experiments using the DIAS dataset. Table~\ref{tab:2} compares the proposed PGM-UNet with several SOTA methods. PGM-UNet achieves outstanding results across multiple metrics, including a DSC of 75.31$\%$, an mIoU of 60.60$\%$, an Acc of 96.20$\%$, an AUC of 87.54$\%$, and a Sen of 77.21$\%$. 

Retinal vessel segmentation plays a crucial role in the diagnosis of various ophthalmic diseases. However, due to the intricate morphology of retinal vessels, it remains a challenging task. To assess the performance of PGM-UNet in vascular segmentation tasks for retinal images, we conduct experiments using the DRIVE. Compared to AttentionUNet, PGM-UNet achieves improvements of 0.41$\%$, 0.57$\%$, 0.05$\%$, 0.38$\%$, and 0.77$\%$ in DSC, mIoU, Acc, AUC, and Sen, respectively. As shown in Table~\ref{tab:3}, PGM-UNet not only outperforms other SOTA methods in overall performance (such as DSC, mIoU, Acc, and AUC) but also excels at capturing finer vascular details, as reflected by its higher Sen.

Fig.~\ref{Fig:4} presents the qualitative results of PGM-UNet on the DIAS and DRIVE. As shown in Fig.~\ref{Fig:4}, PGM-UNet excels at identifying small vessels while maintaining stronger connectivity and smoothness in larger vessels across both DSA and retinal images in the DIAS and DRIVE.
\subsubsection{Generalization Evaluation}
To evaluate the generalization performance of the proposed PGM-UNet, we conduct a generalization experiment in which the model is trained on one dataset and tested on another. This evaluation method more closely reflects real-world clinical applications, making it more clinically significant. In clinical practice, scenarios involving different devices and hospitals are common, necessitating robust and highly generalizable methods. Table~\ref{tab:4} presents the generalization performance of PGM-UNet when trained on ISIC-2018 and tested on PH2. Despite differences in imaging protocols between the ISIC-2018 and PH2, PGM-UNet still achieves strong results across multiple metrics, including a DSC of 92.01$\%$, an mIoU of 85.99$\%$. This fully demonstrates its superior generalization ability compared to other SOTA methods. 
\subsection{Ablation Studies}
To evaluate the contribution of each component in PGM-UNet, we conduct comprehensive ablation studies, systematically assessing the impact of individual modules by removing them sequentially from the architecture.
\subsubsection{The Impact of Multi-Scale Information Extraction Module}
Recognizing the crucial role of the bottleneck layer in connecting the encoder and decoder, we design the MIEM using KANs. To assess its impact, we remove this module from PGM-UNet, creating PGM-UNet w/o MIEM. Table~\ref{tab:5} presents the quantitative results of our ablation study, demonstrating that MIEM significantly improves performance across all metrics.
\subsubsection{The Impact of Local-Global Information Fusion Network}
To assess the contribution of each component in LG-Net, we selectively remove the corresponding parts, generating PGM-UNet w/o LIEM, PGM-UNet w/o PGRM, PGM-UNet w/o Prompt, and PGM-UNet w/o Attention. The corresponding quantitative results are summarized in Table~\ref{tab:5}. The comparison among PGM-UNet w/o LIEM, PGM-UNet w/o PGRM, and PGM-UNet demonstrates that the integration of local and global information significantly improves segmentation performance. Compared to PGM-UNet w/o Prompt, PGM-UNet achieves the optimal performance, primarily due to the effectiveness of the prompt strategy. Additionally, compared to PGM-UNet w/o Attention, PGM-UNet achieves better performance, largely attributable to the channel attention mechanism of the MAFM. Specifically, the channel attention mechanism reweights local and global information, effectively addressing performance degradation caused by information redundancy.
\section{Discussion}
In this study, we have demonstrated that combining CNNs and Mamba in parallel improves segmentation performance in medical image tasks compared to using either approach alone. However, integrating local and global information may introduce redundancy, which degrades model performance. To address this, we employ MAFM with a channel attention mechanism to mitigate redundancy. The channel attention mechanism assigns different weights to local and global information channels, emphasizing the most relevant ones while reducing the influence of irrelevant or redundant channels. As shown in Table~\ref{tab:5}, applying channel attention to reweight these channels before fusion significantly improves medical image segmentation accuracy. In future work, we plan to explore other attention mechanisms to further evaluate the effectiveness of the proposed approach.

Another key contribution of this study is leveraging original input data as visual prompts to enhance global information extraction by PGRM. This visual prompting optimizes information extraction at each stage of the UNet architecture, significantly improving the model’s ability to perceive and process various types of original input data. As shown in Table~\ref{tab:5}, incorporating prompt information leads to a notable improvement in model performance. Thus far, our focus has been on optimizing performance for individual segmentation tasks rather than developing a unified model leveraging adaptive prompting. In future work, we aim to explore additional segmentation tasks and develop an all-in-one medical image segmentation model based on prompt learning.

Finally, this study introduces a parallel dual-path CNN-Mamba architecture to enhance medical image segmentation by enabling more comprehensive information extraction. Although PGM-UNet outperforms SOTA methods, certain limitations persist in skin lesion and vessel segmentation. As shown in Fig.~\ref{Fig:3}, PGM-UNet struggles with skin lesion segmentation when extensive hair occlusion is present. The reliance on global contextual information makes it difficult to accurately distinguish lesion regions under strong hair interference (top row), potentially limiting its effectiveness in clinical applications. Furthermore, Fig.~\ref{Fig:4} highlights challenges in vessel segmentation, particularly in capturing the fine details of complex capillary networks. PGM-UNet shows limitations in segmenting densely distributed and intricate vascular structures, indicating areas for further refinement. Addressing these challenges is crucial for optimizing PGM-UNet for real-world deployment.
\section{Conclusion}
In this paper, we proposed a prompt-guided CNN-Mamba dual-path UNet for medical image segmentation, which effectively integrates both local and global information. The proposed PGRM could adaptively extract dynamic visual prompts from the original input data, effectively guiding Mamba in capturing global information. Additionally, the proposed MIEM enhances contextual feature extraction without altering the resolution. Experimental results demonstrate that PGM-UNet outperforms SOTA methods in medical image segmentation tasks. We conducted cross-dataset experiments to evaluate the generalization capability of PGM-UNet, and the results confirm that our method achieve competitive performance in clinical settings.

\section*{References}
\bibliographystyle{IEEEtran}
\bibliography{ref}

\begin{thebibliography}{10}
\providecommand{\url}[1]{#1}
\csname url@samestyle\endcsname
\providecommand{\newblock}{\relax}
\providecommand{\bibinfo}[2]{#2}
\providecommand{\BIBentrySTDinterwordspacing}{\spaceskip=0pt\relax}
\providecommand{\BIBentryALTinterwordstretchfactor}{4}
\providecommand{\BIBentryALTinterwordspacing}{\spaceskip=\fontdimen2\font plus
\BIBentryALTinterwordstretchfactor\fontdimen3\font minus
  \fontdimen4\font\relax}
\providecommand{\BIBforeignlanguage}[2]{{%
\expandafter\ifx\csname l@#1\endcsname\relax
\typeout{** WARNING: IEEEtran.bst: No hyphenation pattern has been}%
\typeout{** loaded for the language `#1'. Using the pattern for}%
\typeout{** the default language instead.}%
\else
\language=\csname l@#1\endcsname
\fi
#2}}
\providecommand{\BIBdecl}{\relax}
\BIBdecl

\bibitem{Ruan2024VMUNetVM}
\BIBentryALTinterwordspacing
J.~Ruan and S.~Xiang, ``{VM-UNet}: Vision mamba {UNet} for medical image
  segmentation,'' \emph{ArXiv}, vol. abs/2402.02491, 2024. [Online]. Available:
  \url{https://api.semanticscholar.org/CorpusID:267413263}
\BIBentrySTDinterwordspacing

\bibitem{978-981-97-5128-0}
M.~Zhang, Y.~Yu, S.~Jin, L.~Gu, T.~Ling, and X.~Tao, ``{VM-UNET-V2}: Rethinking
  vision mamba {UNet} for medical image segmentation,'' in \emph{Proceedings
  of the Bioinformatics Research and Applications}, 2024, pp. 335--346.

\bibitem{10733833}
F.~Yuan, Y.~Peng, Q.~Huang, and X.~Li, ``A bi-directionally fused boundary
  aware network for skin lesion segmentation,'' \emph{IEEE Transactions on
  Image Processing}, vol.~33, pp. 6340--6353, 2024.

\bibitem{LIU2024103247}
W.~Liu, T.~Tian, L.~Wang, W.~Xu, L.~Li, H.~Li, W.~Zhao, S.~Tian, X.~Pan,
  Y.~Deng, F.~Gao, H.~Yang, X.~Wang, and R.~Su, ``{DIAS}: A dataset and
  benchmark for intracranial artery segmentation in {DSA} sequences,''
  \emph{Medical Image Analysis}, vol.~97, p. 103247, 2024.

\bibitem{MENG2020123}
C.~Meng, K.~Sun, S.~Guan, Q.~Wang, R.~Zong, and L.~Liu, ``Multiscale dense
  convolutional neural network for dsa cerebrovascular segmentation,''
  \emph{Neurocomputing}, vol. 373, pp. 123--134, 2020.

\bibitem{9994763}
X.~Huang, Z.~Deng, D.~Li, X.~Yuan, and Y.~Fu, ``Missformer: An effective
  transformer for 2d medical image segmentation,'' \emph{IEEE Transactions on
  Medical Imaging}, vol.~42, no.~5, pp. 1484--1494, 2023.

\bibitem{7055281}
Y.~Zhao, L.~Rada, K.~Chen, S.~P. Harding, and Y.~Zheng, ``Automated vessel
  segmentation using infinite perimeter active contour model with hybrid region
  information with application to retinal images,'' \emph{IEEE Transactions on
  Medical Imaging}, vol.~34, no.~9, pp. 1797--1807, 2015.

\bibitem{AZZOPARDI201546}
G.~Azzopardi, N.~Strisciuglio, M.~Vento, and N.~Petkov, ``{Trainable COSFIRE}
  filters for vessel delineation with application to retinal images,''
  \emph{Medical Image Analysis}, vol.~19, no.~1, pp. 46--57, 2015.

\bibitem{932744}
Y.~Lee, T.~Hara, H.~Fujita, S.~Itoh, and T.~Ishigaki, ``Automated detection of
  pulmonary nodules in helical {CT} images based on an improved
  template-matching technique,'' \emph{IEEE Transactions on Medical Imaging},
  vol.~20, no.~7, pp. 595--604, 2001.

\bibitem{10.1007/978-3-319-24574-4_28}
O.~Ronneberger, P.~Fischer, and T.~Brox, ``{U-Net}: Convolutional networks for
  biomedical image segmentation,'' in \emph{Proceedings of the Medical Image
  Computing and Computer-Assisted Intervention (MICCAI)}, 2015, pp. 1--8.

\bibitem{Oktay2018AttentionUL}
\BIBentryALTinterwordspacing
O.~Oktay, J.~Schlemper, L.~L. Folgoc, M.~J. Lee, M.~P. Heinrich, K.~Misawa,
  K.~Mori, S.~G. McDonagh, N.~Y. Hammerla, B.~Kainz, B.~Glocker, and
  D.~Rueckert, ``{Attention U-Net}: Learning where to look for the pancreas,''
  \emph{ArXiv}, vol. abs/1804.03999, 2018. [Online]. Available:
  \url{https://api.semanticscholar.org/CorpusID:4861068}
\BIBentrySTDinterwordspacing

\bibitem{10.1007/978-3-030-00889-5_1}
Z.~Zhou, M.~M. Rahman~Siddiquee, N.~Tajbakhsh, and J.~Liang, ``{UNet++}: A
  nested {U-Net} achitecture for medical image segmentation,'' in
  \emph{Proceedings of the Deep Learning in Medical Image Analysis and
  Multimodal Learning for Clinical Decision Support}, 2018, pp. 3--11.

\bibitem{9995040}
J.~Ruan, S.~Xiang, M.~Xie, T.~Liu, and Y.~Fu, ``{MALUNet}: A multi-attention
  and light-weight {UNet} for skin lesion segmentation,'' in \emph{Proceedings
  of the IEEE International Conference on Bioinformatics and Biomedicine
  (BIBM)}, 2022, pp. 1150--1156.

\bibitem{Chen2021TransUNetTM}
\BIBentryALTinterwordspacing
J.~Chen, Y.~Lu, Q.~Yu, X.~Luo, E.~Adeli, Y.~Wang, L.~Lu, A.~L. Yuille, and
  Y.~Zhou, ``{TransUNet}: Transformers make strong encoders for medical image
  segmentation,'' \emph{ArXiv}, vol. abs/2102.04306, 2021. [Online]. Available:
  \url{https://api.semanticscholar.org/CorpusID:231847326}
\BIBentrySTDinterwordspacing

\bibitem{Dosovitskiy2020AnII}
\BIBentryALTinterwordspacing
A.~Dosovitskiy, L.~Beyer, A.~Kolesnikov, D.~Weissenborn, X.~Zhai,
  T.~Unterthiner, M.~Dehghani, M.~Minderer, G.~Heigold, S.~Gelly, J.~Uszkoreit,
  and N.~Houlsby, ``An image is worth 16x16 words: Transformers for image
  recognition at scale,'' \emph{ArXiv}, vol. abs/2010.11929, 2020. [Online].
  Available: \url{https://api.semanticscholar.org/CorpusID:225039882}
\BIBentrySTDinterwordspacing

\bibitem{10.1007/978-3-031-25066-8_9}
H.~Cao, Y.~Wang, J.~Chen, D.~Jiang, X.~Zhang, Q.~Tian, and M.~Wang,
  ``{Swin-Unet}: Unet-like pure transformer for medical image segmentation,''
  in \emph{Proceedings of the European Conference on Computer Vision (ECCV)},
  2023, pp. 205--218.

\bibitem{Liu2021SwinTH}
\BIBentryALTinterwordspacing
Z.~Liu, Y.~Lin, Y.~Cao, H.~Hu, Y.~Wei, Z.~Zhang, S.~Lin, and B.~Guo, ``Swin
  transformer: Hierarchical vision transformer using shifted windows,''
  \emph{Proceedings of the IEEE/CVF International Conference on Computer Vision
  (ICCV)}, pp. 9992--10\,002, 2021. [Online]. Available:
  \url{https://api.semanticscholar.org/CorpusID:232352874}
\BIBentrySTDinterwordspacing

\bibitem{10.1007/978-3-030-87193-2_2}
Y.~Zhang and H.~Liu, ``{TransFuse}: Fusing transformers and {CNN}s for medical
  image segmentation,'' in \emph{Proceedings of the Medical Image Computing and
  Computer Assisted Intervention (MICCAI)}, 2021, pp. 14--24.

\bibitem{Gu2021EfficientlyML}
\BIBentryALTinterwordspacing
A.~Gu, K.~Goel, and C.~R'e, ``Efficiently modeling long sequences with
  structured state spaces,'' \emph{ArXiv}, vol. abs/2111.00396, 2021. [Online].
  Available: \url{https://api.semanticscholar.org/CorpusID:240354066}
\BIBentrySTDinterwordspacing

\bibitem{Gu2023MambaLS}
\BIBentryALTinterwordspacing
A.~Gu and T.~Dao, ``Mamba: Linear-time sequence modeling with selective state
  spaces,'' \emph{ArXiv}, vol. abs/2312.00752, 2023. [Online]. Available:
  \url{https://api.semanticscholar.org/CorpusID:265551773}
\BIBentrySTDinterwordspacing

\bibitem{Ma2024UMambaEL}
\BIBentryALTinterwordspacing
J.~Ma, F.~Li, and B.~Wang, ``{U-Mamba}: Enhancing long-range dependency for
  biomedical image segmentation,'' \emph{ArXiv}, vol. abs/2401.04722, 2024.
  [Online]. Available: \url{https://api.semanticscholar.org/CorpusID:266899624}
\BIBentrySTDinterwordspacing

\bibitem{978-3-031-72111-3}
Z.~Xing, T.~Ye, Y.~Yang, G.~Liu, and L.~Zhu, ``{SegMamba}: Long-range
  sequential modeling mamba for {3D} medical image segmentation,'' in
  \emph{Proceedings of the Medical Image Computing and Computer Assisted
  Intervention (MICCAI)}, 2024, pp. 578--588.

\bibitem{Liu2024VMambaVS}
\BIBentryALTinterwordspacing
Y.~Liu, Y.~Tian, Y.~Zhao, H.~Yu, L.~Xie, Y.~Wang, Q.~Ye, and Y.~Liu,
  ``{VMamba}: Visual state space model,'' \emph{ArXiv}, vol. abs/2401.10166,
  2024. [Online]. Available:
  \url{https://api.semanticscholar.org/CorpusID:267035250}
\BIBentrySTDinterwordspacing

\bibitem{Liao2024LightMUNetMA}
\BIBentryALTinterwordspacing
W.~Liao, Y.~Zhu, X.~Wang, C.~Pan, Y.~Wang, and L.~Ma, ``{LightM-UNet}: Mamba
  assists in lightweight {UNet} for medical image segmentation,'' \emph{ArXiv},
  vol. abs/2403.05246, 2024. [Online]. Available:
  \url{https://api.semanticscholar.org/CorpusID:268297157}
\BIBentrySTDinterwordspacing

\bibitem{10.1007/978-3-031-76197-3_2}
V.-T. Nguyen, V.-T. Pham, and T.-T. Tran, ``{AC-MambaSeg}: An adaptive
  convolution and mamba-based architecture for enhanced skin lesion
  segmentation,'' in \emph{Computational Intelligence Methods for Green
  Technology and Sustainable Development}, 2024, pp. 13--26.

\bibitem{10.1007/978-3-030-01234-2_18}
Y.~Zhang, K.~Li, K.~Li, L.~Wang, B.~Zhong, and Y.~Fu, ``Image super-resolution
  using very deep residual channel attention networks,'' in \emph{Proceedings
  of the European Conference on Computer Vision (ECCV)}, 2018, pp. 294--310.

\bibitem{potlapalli2023promptir}
V.~Potlapalli, S.~W. Zamir, S.~Khan, and F.~Khan, ``Promptir: Prompting for
  all-in-one image restoration,'' in \emph{Proceedings of the Conference on
  Neural Information Processing Systems (NeurIPS)}, 2023.

\bibitem{10833823}
X.~Xie, W.~Zhao, M.~Nan, Z.~Zhang, Y.~Wu, H.~Zheng, D.~Liang, M.~Wang, and
  Z.~Hu, ``Prompt-agent-driven integration of foundation model priors for
  low-count pet reconstruction,'' \emph{IEEE Transactions on Medical Imaging},
  pp. 1--1, 2025.

\bibitem{Kong2024TowardsEM}
\BIBentryALTinterwordspacing
X.~Kong, C.~Dong, and L.~Zhang, ``Towards effective multiple-in-one image
  restoration: A sequential and prompt learning strategy,'' \emph{ArXiv}, vol.
  abs/2401.03379, 2024. [Online]. Available:
  \url{https://api.semanticscholar.org/CorpusID:266844752}
\BIBentrySTDinterwordspacing

\bibitem{1957On}
A.~N. Kolmogorov, ``On the representation of continuous functions of many
  variables by superposition of continuous functions of one variable and
  addition,'' \emph{American Mathematical Society Translations}, vol.~28, pp.
  55--59, 1957.

\bibitem{Liu2024KANKN}
\BIBentryALTinterwordspacing
Z.~Liu, Y.~Wang, S.~Vaidya, F.~Ruehle, J.~Halverson, M.~Soljacic, T.~Y. Hou,
  and M.~Tegmark, ``{KAN}: Kolmogorov-arnold networks,'' \emph{ArXiv}, vol.
  abs/2404.19756, 2024. [Online]. Available:
  \url{https://api.semanticscholar.org/CorpusID:269457619}
\BIBentrySTDinterwordspacing

\bibitem{Li2024UKANMS}
\BIBentryALTinterwordspacing
C.~Li, X.~Liu, W.~Li, C.~Wang, H.~Liu, and Y.~Yuan, ``{U-KAN} makes strong
  backbone for medical image segmentation and generation,'' \emph{ArXiv}, vol.
  abs/2406.02918, 2024. [Online]. Available:
  \url{https://api.semanticscholar.org/CorpusID:270258333}
\BIBentrySTDinterwordspacing

\bibitem{Li2024HSRKANEH}
\BIBentryALTinterwordspacing
B.~Li, X.~Wang, and H.~Xu, ``{HSR-KAN}: Efficient hyperspectral image
  super-resolution via kolmogorov-arnold networks,'' \emph{ArXiv}, vol.
  abs/2409.06705, 2024. [Online]. Available:
  \url{https://api.semanticscholar.org/CorpusID:272593269}
\BIBentrySTDinterwordspacing

\bibitem{8662594}
Z.~Gu, J.~Cheng, H.~Fu, K.~Zhou, H.~Hao, Y.~Zhao, T.~Zhang, S.~Gao, and J.~Liu,
  ``{CE-Net}: Context encoder network for {2D} medical image segmentation,''
  \emph{IEEE Transactions on Medical Imaging}, vol.~38, no.~10, pp. 2281--2292,
  2019.

\bibitem{9662196}
A.~Srivastava, D.~Jha, S.~Chanda, U.~Pal, H.~D. Johansen, D.~Johansen, M.~A.
  Riegler, S.~Ali, and P.~Halvorsen, ``{MSRF-Net}: A multi-scale residual
  fusion network for biomedical image segmentation,'' \emph{IEEE Journal of
  Biomedical and Health Informatics}, vol.~26, no.~5, pp. 2252--2263, 2022.

\bibitem{XU2023106626}
Q.~Xu, Z.~Ma, N.~He, and W.~Duan, ``{DCSAU-Net}: A deeper and more compact
  split-attention {U-Net} for medical image segmentation,'' \emph{Computers in
  Biology and Medicine}, vol. 154, p. 106626, 2023.

\bibitem{8363547}
N.~C.~F. Codella, D.~Gutman, M.~E. Celebi, B.~Helba, M.~A. Marchetti, S.~W.
  Dusza, A.~Kalloo, K.~Liopyris, N.~Mishra, H.~Kittler, and A.~Halpern, ``Skin
  lesion analysis toward melanoma detection: A challenge at the 2017
  international symposium on biomedical imaging (isbi), hosted by the
  international skin imaging collaboration (isic),'' in \emph{Proceedings of
  the IEEE International Symposium on Biomedical Imaging (ISBI)}, 2018, pp.
  168--172.

\bibitem{Codella2019SkinLA}
\BIBentryALTinterwordspacing
N.~C.~F. Codella, V.~M. Rotemberg, P.~Tschandl, M.~E. Celebi, S.~W. Dusza,
  D.~Gutman, B.~Helba, A.~Kalloo, K.~Liopyris, M.~A. Marchetti, H.~Kittler, and
  A.~C. Halpern, ``Skin lesion analysis toward melanoma detection 2018: A
  challenge hosted by the international skin imaging collaboration (isic),''
  \emph{ArXiv}, vol. abs/1902.03368, 2019. [Online]. Available:
  \url{https://api.semanticscholar.org/CorpusID:60440592}
\BIBentrySTDinterwordspacing

\bibitem{tschandl2018ham10000}
P.~Tschandl, C.~Rosendahl, and H.~Kittler, ``The ham10000 dataset, a large
  collection of multi-source dermatoscopic images of common pigmented skin
  lesions,'' \emph{Sci. Data}, vol.~5, 2018.

\bibitem{1282003}
J.~Staal, M.~Abramoff, M.~Niemeijer, M.~Viergever, and B.~van Ginneken,
  ``Ridge-based vessel segmentation in color images of the retina,'' \emph{IEEE
  Transactions on Medical Imaging}, vol.~23, no.~4, pp. 501--509, 2004.

\bibitem{6610779}
T.~Mendonça, P.~M. Ferreira, J.~S. Marques, A.~R.~S. Marcal, and J.~Rozeira,
  ``{PH2} - a dermoscopic image database for research and benchmarking,'' in
  \emph{2013 35th Annual International Conference of the IEEE Engineering in
  Medicine and Biology Society (EMBC)}, 2013, pp. 5437--5440.

\end{thebibliography}
\end{document}